\documentclass{aa}
\newcommand{\varcsec}{^{\prime\prime}}
\newcommand{\arcsecc}{.\hspace{-0.9mm}'\!\hskip0.4pt'\hspace{-0.2mm}}
\newcommand\ionn[2]{#1$\;${\scshape{#2}}}% 

\usepackage{amssymb}
\usepackage{textcomp}
\usepackage{epstopdf}
\usepackage{graphicx}
\usepackage{float}
\usepackage{microtype}
\usepackage{color} 
\usepackage{xfrac} 
\usepackage{hyphenat}
\usepackage{inputenc}
\usepackage{fontenc}
\usepackage{natbib}
\usepackage{csquotes}
\usepackage{txfonts}
\usepackage[bookmarks = true,colorlinks = true,linkcolor = blue,urlcolor = blue,citecolor = blue,bookmarks = true,linktocpage = true,hyperindex = true,breaklinks]{hyperref}
\bibpunct{(}{)}{;}{a}{}{,}

\begin{document}
   \title{Upper Chromospheric Magnetic Field of a Sunspot Penumbra: Observations of Fine Structure}
   
   \subtitle{}

  \author{J. Joshi\inst{\ref{ISP},\ref{MPS}}
           \and A. Lagg \inst{\ref{MPS}}
           \and S. K. Solanki\inst{\ref{MPS},\ref{SSR}}
           \and A. Feller\inst{\ref{MPS}}
           \and M. Collados\inst{\ref{IAC},\ref{DAUL}}
           \and D. Orozco Su\'{a}rez\inst{\ref{IAC},\ref{DAUL}}
           \and R. Schlichenmaier\inst{\ref{KIS}} 
           \and M. Franz\inst{\ref{KIS}}
           \and H. Balthasar\inst{\ref{LIAP}}
           \and C. Denker\inst{\ref{LIAP}}
           \and T. Berkefeld\inst{\ref{KIS}}
           \and A. Hofmann\inst{\ref{LIAP}}
           \and C. Kiess\inst{\ref{KIS}}
           \and H. Nicklas\inst{\ref{IAGU}}
           \and A. Pastor Yabar\inst{\ref{IAC},\ref{DAUL}}
           \and R. Rezaei\inst{\ref{IAC},\ref{DAUL}}      
           \and D. Schmidt\inst{\ref{KIS}}
           \and W. Schmidt\inst{\ref{KIS}}
           \and M. Sobotka\inst{\ref{AIAS}}
           \and D. Soltau\inst{\ref{KIS}}
           \and J. Staude\inst{\ref{LIAP}}
           \and K. G. Strassmeier\inst{\ref{LIAP}}
           \and R. Volkmer\inst{\ref{KIS}}
           \and O. von der L\"{u}he\inst{\ref{KIS}}
           \and T. Waldmann\inst{\ref{KIS}}}           
   \institute{Institute for Solar Physics, Department of Astronomy, Stockholm University, AlbaNova University Centre,
              SE-106 91 Stockholm, Sweden, \label{ISP} \email{jayant.joshi@astro.su.se} 
              \and  Max-Planck-Institut f\"{u}r Sonnensystemforschung, Justus-von-Liebig-Weg 3,
              37077, G\"{o}ttingen, Germany\label{MPS} 
              \and School of Space Research, Kyung Hee University, Yongin, Gyeonggi Do, 446-701, 
              Republic of Korea\label{SSR} 
              \and Instituto de Astrof\'{\i}sica de Canarias (IAC), V\'{\i}a Lact\'{e}a, 38200 La Laguna, Tenerife, Spain\label{IAC}
              \and Departamento de Astrof\'{\i}sica, Universidad de La Laguna, 38205 La Laguna, Tenerife, Spain\label{DAUL}
              \and Kiepenheuer-Institut f\"{u}r Sonnenphysik, Sch\"{o}neckstr. 6, 79104 Freiburg, Germany\label{KIS}
              \and Leibniz Institut f\"{u}r Astrophysik Potsdam (AIP), An der Sternwarte 16, 14482 Potsdam, Germany\label{LIAP}
              \and Institut f\"{u}r Astrophysik, Georg-August-Universit\"{a}t G\"{o}ttingen, Friedrich-Hund-Platz 1, 37077 G\"{o}ttingen, Germany\label{IAGU}
              \and Astronomical Institute of the Academy of Sciences, Fri\v{c}ova 298, 25165 Ond\v{r}ejov, Czech Republic\label{AIAS}
              }
   \date{Received; accepted}

   \titlerunning{Penumbral fine structure in the upper chromosphere}
   \authorrunning{Jayant Joshi et. al.}

% \abstract{}{}{}{}{} 
% 5 {} token are mandatory

\abstract
  {}  
 {The fine-structure of magnetic field of a sunspot penumbra in the upper chromosphere is to be explored and compared
 to that in the photosphere.}
 {High spatial resolution spectropolarimetric observations were recorded with the 1.5-meter GREGOR telescope using the
 GREGOR Infrared Spectrograph (GRIS). The observed spectral domain includes
 the upper chromospheric \ionn{He}{i}\, triplet at 10830\,\AA\, and the photospheric \ionn{Si}{i}\,10827.1\,\AA\, and 
 \ionn{Ca}{i}\,10833.4\,\AA\, spectral lines. The upper chromospheric magnetic field is obtained by inverting the 
 \ionn{He}{i}\, triplet assuming a Milne-Eddington type model atmosphere. A height dependent inversion was applied to
 the \ionn{Si}{i}\,10827.1\,\AA\, and \ionn{Ca}{i}\,10833.4\,\AA\, lines to obtain the photospheric magnetic field.}
 {We find that the inclination of the magnetic field shows variations in the azimuthal direction both in the photosphere, 
 but also in the upper chromosphere. The chromospheric variations remarkably well coincide with the variations in the 
 inclination of the photospheric field and resemble the well-known spine and inter-spine structure in the photospheric 
 layers of penumbrae. The typical peak-to-peak variations in the inclination of the magnetic field in the upper 
 chromosphere is found to be $10^\circ$--$15^\circ$, i.e., roughly half the variation in the photosphere. In contrast, the
 magnetic field strength of the observed penumbra does not show variations on small spatial scales in the upper chromosphere. 
 }
   {Thanks to the high spatial resolution observations possible with the GREGOR telescope at 1.08 microns,
   we find that the prominent small-scale fluctuations in the magnetic field inclination, which are a salient
   part of the property of sunspot penumbral photospheres, also persist in the chromosphere, although at somewhat 
   reduced amplitudes. Such a complex magnetic configuration may facilitate penumbral chromospheric dynamic
   phenomena, such as penumbral micro-jets or transient bright dots.}


\maketitle
%
%________________________________________________________________

\section{Introduction} \label{sec_1}

In photospheric layers, the magnetic field of sunspot penumbrae possesses a highly complex structure, varying in both,
strength and orientation on fine spatial scales. It is organized in nearly radially aligned filaments (at least in the 
inner penumbra) where strong and weak magnetic fields are alternatingly present in azimuthal direction. Similarly, the
magnetic field orientation is organized in vertical (spines) and horizontal (inter-spines) components 
\citep{Schmidt_1992,Title_1993,Solanki_1993a,Martinez_2000a,Langhans_2005,Scharmer_2012,Tiwari_2013}. Relations between
various physical parameters of penumbrae, such as the magnetic field strength, field orientation, brightness and plasma 
flows (Evershed effect), have been studied for the last few decades, leading to some controversial results \citep[for 
detailed reviews on the topic see, e.g.,][]{Solanki_2003, Borrero_2011,Rempel_2011b}. A recent study by \citet{Tiwari_2013}
seems to settle most of the controversies and presented a uniform picture of penumbral filaments. They conclude that the
penumbral filaments harbor a comparatively weak horizontal magnetic field except at its head, where the field is relatively
vertical, and at tail, where the field has opposite polarity. Such filaments are, at least in the inner part of the 
penumbra embedded in strong and less inclined background magnetic field.

Another feature of the photospheric penumbral magnetic field is the presence of opposite polarity patches (compared to
the polarity of the umbra) at the tails and edges of penumbral filaments \citep{Scharmer_2013,Ruiz_2013,Franz_2013,
Tiwari_2013,Vannoort_2013,Joshi_2014,Joshi_2016a}. These opposite polarity patches are thought to originate from advection
of penumbral field by overturning convection. The presence of overturning convection in penumbrae has been successfully 
observed \citep{Joshi_2011,Scharmer_2011,Tiwari_2013,Esteban_2015}.

The chromosphere and transition region above the sunspot penumbrae display dynamics on arcsec and sub-arcsec scales, 
such as penumbral micro-jets \citep{2007Sci...318.1594K, 2008A&A...488L..33J, 2013ApJ...779..143R, 2015ApJ...811L..33V, 
2016ApJ...816...92T} and transient bright dots \citep{2014ApJ...790L..29T}. \citet{2007Sci...318.1594K} suggest reconnection
between two magnetic components (spines and inter-spines) as a driver of penumbra jets, whereas \citet{2016ApJ...816...92T}
argued for reconnection between opposite polarity patches and fields from spines \citep[cf.][]{2010ApJ...715L..40M}. 
\cite{2014ApJ...790L..29T} also speculate that bright dots they observed in the penumbra may originate from heating of
plasma to transition region temperature due to reconnection of magnetic fields from spine and inter-spine regions. In 
literature small-scale transient events in the chromosphere and transition region above sunspot penumbrae are usually 
linked to the complex magnetic structure in the photosphere. Does the chromospheric magnetic field of penumbrae play 
any role in these dynamics? This topic has not been investigated because information about the magnetic field fine 
structure of sunspots at chromospheric heights is lagging behind that in the photosphere, especially in terms of spatial
resolution. The general structure of sunspot magnetic fields has been studied by \citet{Rueedi_1995b}, \citet{Joshi_2014},
\citet{Schad_2015} and \citet{Joshi_2016b} by using observation in the \ionn{He}{i}\, triplet at 10830\,\AA.  

In this paper, we present the first high spatial resolution maps of the penumbral magnetic field in the upper chromosphere
obtained from the \ionn{He}{i}\, triplet at 10830\,\AA\ observed with the 1.5-meter GREGOR telescope.

\begin{figure}
\centering
      \includegraphics[width = 0.48\textwidth]{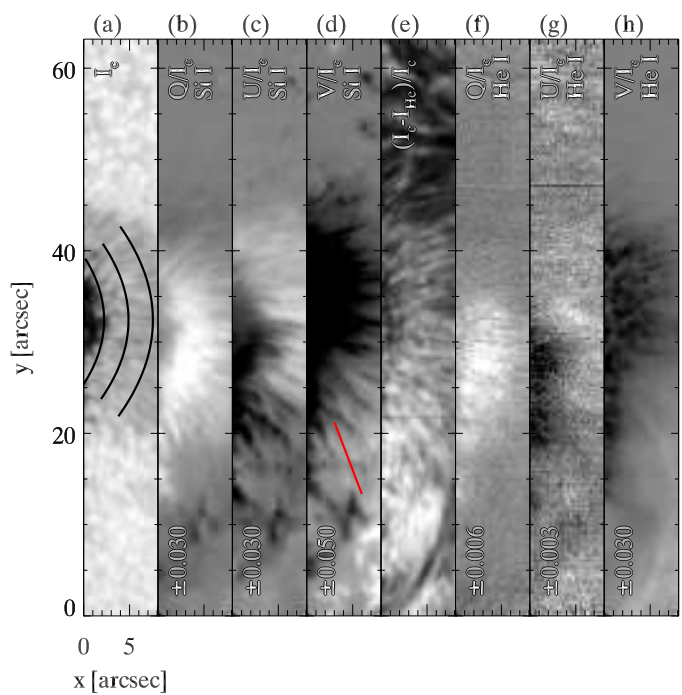}
      \caption{Maps of Stokes parameters for the \ionn{Si}{i}\,10827.1\,\AA\, line and the \ionn{He}{i}\, triplet at 
      10830\,\AA. Panel (a) shows continuum intensity, $I_{\rm{c}}$, at 10833\,\AA. Panels (b), (c) and (d) corresponds
      to Stokes $Q/I_{c}$, $U/I_{c}$ and $V/I_{c}$ maps of the \ionn{Si}{i}\, line, respectively. The line depression 
      of the blended red components of the \ionn{He}{i}\, triplet, $(I_{\rm{c}}-I_{\rm{He}})/I_{\rm{c}}$, is displayed 
      in panel (e). Stokes $Q/I_{c}$, $U/I_{c}$ and $V/I_{c}$ maps for the \ionn{He}{i}\, triplet are presented in panels
      (e), (f) and (g), respectively. The lower and upper limits of the Stokes $Q/I_{c}$, $U/I_{c}$ and $V/I_{c}$ maps are 
      indicated in the lower part of the respective image. The Stokes maps presented here are an average over a 0.27\,\AA\, 
      wide range centered at $\lambda-\lambda_{\rm{0}} = -$3.10\,\AA\, and $\lambda - \lambda_{\rm{0}} = -$0.35\,\AA\, for
      the \ionn{Si}{i}\, line and \ionn{He}{i}\, triplet, respectively, with $\lambda_{\rm{0}}$ = 10830\,\AA. Stokes 
      $V/I_{c}$ profiles along the red line marked in panel (d) are plotted in Fig.~\ref{prof}. The lower part of the line is 
      directed towards the closest part of the solar limb. Three arcs in panel (a) show the locations of various parameters
      which are depicted in Fig.~\ref{par_prof}.} 
      \label{FOV}
\end{figure}

\begin{figure}
\centering
      \includegraphics[width = 0.47\textwidth]{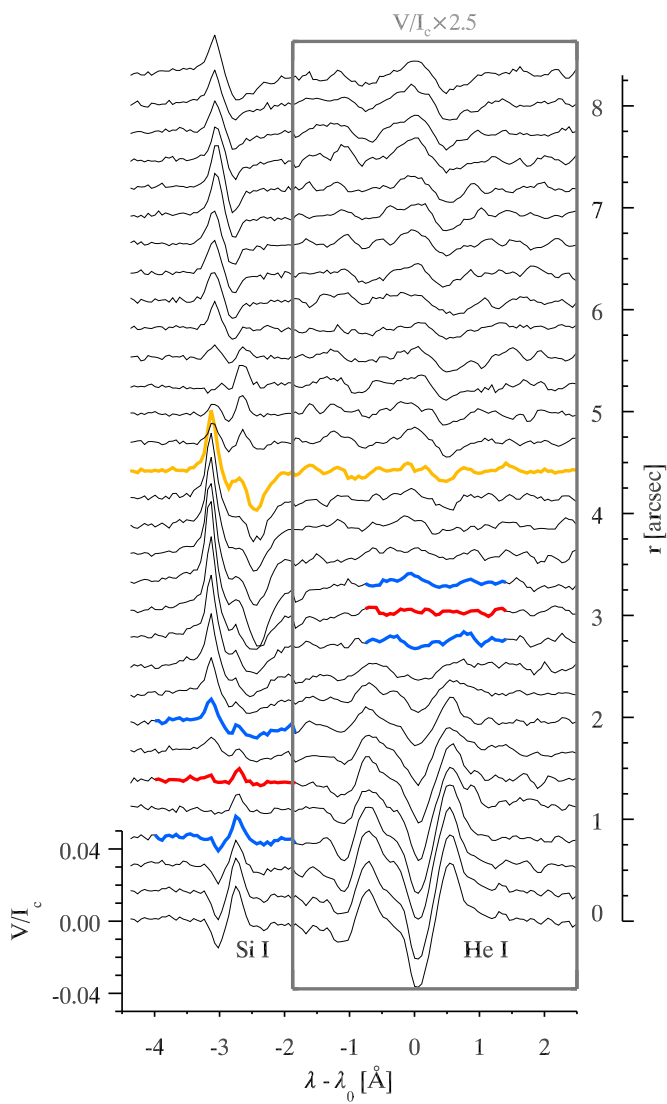}
      \caption{Stokes $V$ profiles over a spectral range covering the \ionn{Si}{i}\,10827.1\,\AA\, line and \ionn{He}{i}\,
      triplet at 10830\,\AA. Positions of the plotted profiles are indicated by a red line in Fig.~\ref{FOV}(d), the lowermost
      profile belongs to the pixel closest within the sunspot to the sunspot center. Since the Stokes $V$ signal in the 
      \ionn{He}{i}\, triplet is weaker compared to the \ionn{Si}{i}\, line, it is magnified by a factor of 2.5 (see box). 
      The axis right side of the plot gives the distance, $r$, in seconds of arc from the lowermost profile. Profiles plotted
      in red indicate the position of the polarity inversion line and those colored in blue indicate the nearest profiles with 
      clearly visible opposite polarity. A yellow profile at $r = 4\arcsecc5$ marks the transition from the penumbral region
      to the quiet Sun.}
      \label{prof}
\end{figure}

\section{Observations} \label{sec_2}

We observed a part of a sunspot penumbra in the active region NOAA 12096  on 25 June 2014 with the 1.5-meter GREGOR 
telescope \citep{2012AN....333..796S}. Spectropolarimetric observations were recorded with the GREGOR 
Infrared Spectrograph \citep[GRIS,][]{2012AN....333..872C} combined with the Tenerife Infrared Polarimeter-2 
\citep[TIP-2,][]{Collados_2007}. The GREGOR telescope is equipped with the GREGOR Adaptive Optics System 
\citep[GAOS,][]{2012AN....333..863B,2010ApOpt..49G.155B} which worked during the observations under 
good seeing conditions. 

We observed a 18\,\AA\, wide range centered around 10833\,\AA\, with a spectral 
sampling of 18~m\AA. The observed field-of-view (FOV) was obtained by a raster scan of the $63\arcsecc3$ 
long slit with 60 steps, and a step size of $0\arcsecc135$. The pixel size along the slit is 
$0\arcsecc135$. Full Stokes profiles were recorded with an exposure time of 40~ms, with ten accumulations 
for every Stokes parameter. The heliocentric coordinates of the centers of the observed FOV were 
($7^\circ$N, $38^\circ$E), so that the cosine of the heliocentric angle is $\mu$ = 0.78.

Standard data reduction steps have been applied to the raw data (see Collados et al., in preparation), which 
encompass procedures such as, corrections for dark current and flat-field, removal of geometrical distortions 
and fringes as well as polarimetric calibration and cross-talk removal. The continuum level was corrected using a Mcath-Pierce Fourier Transform 
Spectrometer (FTS) spectrum \citep{Livingston_1991,Wallace_1993}. The spectral resolution of the recorded spectra 
is estimated in a similar way as in \citet{Borrero_2016} by least-square fitting the FTS spectrum convolved with 
a Gaussian spectral point-spread-function (PSF) to the averaged 
quiet-Sun spectrum (recorded in flat-field mode on the disk center). The best fit between the FTS and averaged quiet Sun 
spectrum is found for a PSF with FWHM = 120~m\AA\, and a spectral straylight of 14\%. This value of FWHM 
corresponds to a spectral resolution of $\lambda/\Delta\lambda\simeq 90,000$, which is similar to that
found by \citet{Borrero_2016} for the GRIS/GREGOR observations recorded at $1.56\,\rm{\mu m}$ \citep[also see,][]{Franz_2016}.

We estimated the spatial resolution of the observations to be $\sim0\arcsecc35$ by looking at the spatial power spectrum. 
For comparison, the diffraction limited resolution of the GREGOR at 10832\,\AA\, is 
$\sim0\arcsecc18$, and the best resolution that GRIS can achieve at this wavelength is $\sim0\arcsecc27$ (twice the pixel size). 
To improve the signal-to-noise ratio of Stokes parameter we binned data by two pixels both in the slit direction and the 
scanning direction (effective spatial resolution $\sim0\arcsecc54$) as well as three pixels in the direction of spectral dispersion.
This implies reducing the spatial resolution somewhat, but greatly enhances the signal-to-noise ratio of the data, allowing for 
much more reliable maps of the magnetic vector in the chromosphere.
Figure~\ref{FOV} presents the observed continuum intensity, $I_{\rm{c}}$, at 10833\,\AA\, over the observed FOV 
(panel (a)), line depression of the blended red components of the \ionn{He}{i} triplet at 10830\,\AA, $(I_{\rm{c}} - I_{\rm{He}})/I_{\rm{c}}$
(panel (e)). Stokes $Q$, $U$ and $V$ maps for the \ionn{Si}{i}\,10827.1\,\AA\, line are shown
in panels (b)--(d) and those for the \ionn{He}{i} triplet are displayed in panels (f)--(h).

\section{Inversions} \label{sec_4}

\subsection{ The \ionn{He}{i}\,10830\,\AA\, triplet}

The \ionn{He}{i}\, triplet at 10830\,\AA\, originates from the transition between the lower $1s2s~^{3}S_{1}$ and the 
upper $1s2p~^{3}P_{0,1,2}$ energy levels, generating three spectral lines: a blue component at 10829.09\,\AA\, 
and two blended red components at 10830.30\,\AA. The \ionn{He}{i}\, triplet is strongly influenced by EUV coronal irradiation, 
capable of ionizing neutral helium atoms, which then, by recombination processes, populate the lower level of the transition
\citep{Avrett_1994,Andretta_1997,Centeno_2008,Leenaarts_2016}.

The \ionn{He}{i}\, triplet is thought to be formed near the top of the chromosphere because the UV radiation does not 
penetrate deep into the chromosphere, populating the ground state of the \ionn{He}{i}\, triplet only in the upper layers.
We inverted the \ionn{He}{i}\, triplet to obtain the upper chromospheric magnetic field and 
the Doppler velocities with the HeLIx$^{\rm{+}}$ \citep{Lagg_2004,Lagg_2009} inversion code assuming a 
Milne-Eddington type model atmosphere which has proved to be a reasonable approximation for the \ionn{He}{i}\, triplet.  
We used eight free parameters in the model atmosphere to
fit the observed Stokes profiles in the \ionn{He}{i}\, triplet: magnetic field strength, 
inclination and azimuth of the magnetic field vector, the Doppler velocity, the Doppler broadening, the damping constant, 
gradient of the source function and the ratio between line-center and the continuum opacity. The \ionn{Si}{i}\,10827.1\,\AA\,
line blends with the blue component of the \ionn{He}{i}\, triplet. To take this blend into account we also fit the \ionn{Si}{i}\,
line, together with the \ionn{He}{i}\, triplet \citep[for more details see][]{Joshi_2016a}.

\begin{figure*}
\centering
       \includegraphics[width = 0.9\textwidth]{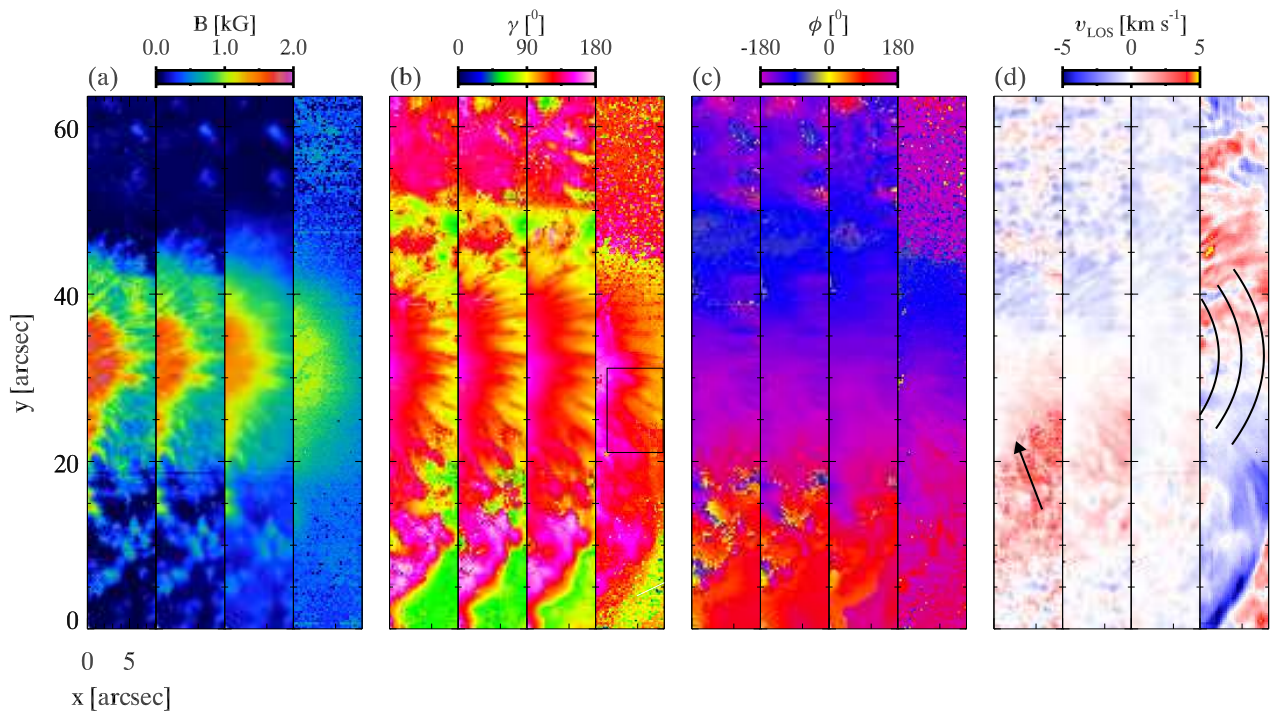}
       \caption{Maps of the magnetic field vector (from left to right: magnetic field strength, $B$, inclination, $\gamma$, 
       azimuth, $\phi$) and LOS-velocity, $\upsilon_{\rm LOS}$, in the photosphere and upper chromosphere. 
       The first three maps in every panel correspond to $\log \tau = 0.0$, $\log \tau =- 0.7$ and $\log \tau = -2.3$ 
       (SPINOR inversion of the \ionn{Si}{i}\, and \ionn{Ca}{i}\, lines), and the fourth map refer to the upper chromosphere
       (HeLIx$^{\rm{+}}$ inversion of the \ionn{He}{i}\, triplet). The box in the fourth map of the panel (b) represents the location for which
       details of various parameters are displayed in Fig.~\ref{par_map_roi}. Three arcs in panel (d) show the locations of 
      various parameters which are depicted in Fig.~\ref{par_prof} and \ref{inc_vel}. An arrow in panel (d) indicates the disk center direction.}     
       \label{par_map}
\end{figure*}

\subsection{ The \ionn{Si}{i}\,10827.1\,\AA\, and \ionn{Ca}{i}\,10833.4\,\AA\, lines}

To retrieve the height-dependent photospheric atmosphere, simultaneous inversions of the \ionn{Si}{i}\,10827.1\,\AA\, and 
\ionn{Ca}{i}\,10833.4\,\AA\, lines are performed to satisfy the radiative transfer equation (RTE) under the 
assumption of local thermodynamic equilibrium (LTE).  We used the SPINOR inversion code \citep{Frutiger_2000}
which is based on the STOPRO routines \citep{Solanki_1987_PhDT}. We used the same set of line 
parameters as used by \citet{Joshi_2016a}.

The model atmosphere used to fit the observed Stokes profiles of the \ionn{Si}{i}\, and \ionn{Ca}{i}\ lines
consists of the temperature, the Doppler velocity, the magnetic field strength, inclination and azimuth as free 
parameters at three nodes, $\log \tau_{\rm{1083}} = 0.0, -0.7, -2.3$. The micro- and macro-turbulence are assumed 
to be constant with height. Here $\tau_{\rm{1083}}$ denotes the optical depth at 1083\,nm. The node positions
are chosen based on the analysis of the magnetic field response function as carried out by \citet{Joshi_2016a} \citep[also see 
Chapt.~4 of][]{Joshi_2014} for the lines under consideration.

We are aware of the fact that the \ionn{Si}{i}\,10827.1\,\AA\, line-core forms under non-local thermodynamic equilibrium
(NLTE) conditions \citep[see][]{Bard_2008}. \citet{2012A&A...539A.131K} have demonstrated that the temperature is the most 
affected atmospheric parameter when NLTE is not taken into account for the synthesis of this line. The 
effect on the magnetic field and Doppler velocities is negligible. In this paper, we only discuss the
magnetic field and Doppler velocities, justifying the assumption of LTE.\\  

\noindent
The spectral PSF (plus spectral straylight, 14\%) of the observations is self-consistently 
included in the SPINOR and HeLIx$^{\rm{+}}$ inversions. We have not included the spatial PSF of the GREGOR and also 
ignored spatial straylight in the inversions. This washes out contrasts, but should not affect our results qualitatively.

\begin{figure}
\centering
      \includegraphics[width = 0.47\textwidth]{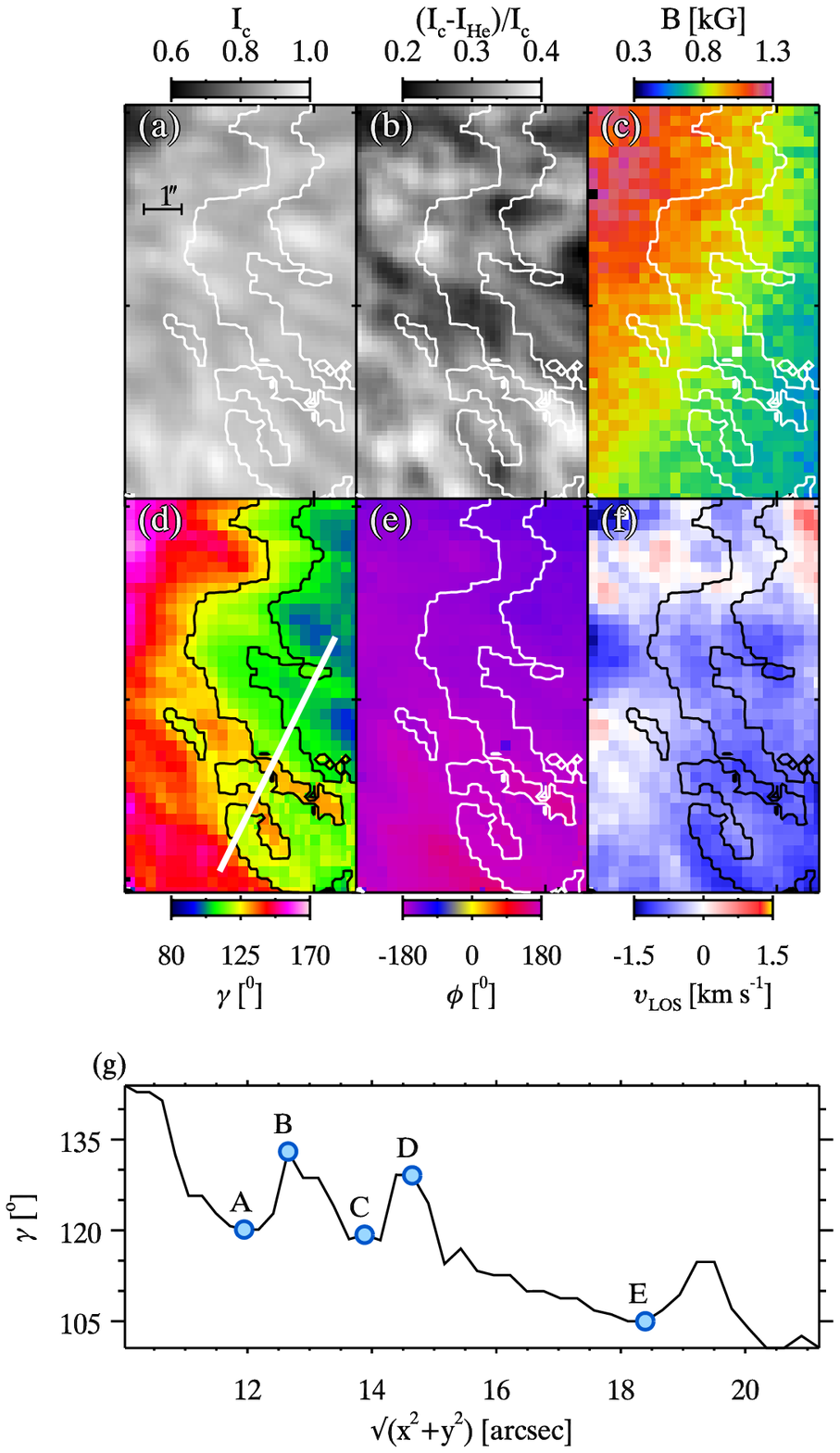}
      \caption{Details of penumbral fine structure in the upper chromosphere. (a) $I_{\rm{c}}$, (b) $(I_{\rm{c}}-I_{\rm{He}})/I_{\rm{c}}$, 
      (c) $B$, (d) $\gamma$, (e) $\phi$ and (f) $\upsilon_{\rm{LOS}}$. The location of the maps is indicated by a box in the
      fourth map of Fig.~\ref{par_map}(b). Contours in panels (a)--(f) depict $\gamma =  110^{\circ}, 125^{\circ}$. Panel (g) depicts variations in $\gamma$ along the white line marked in 
      panel (d). Blue circles identified as A--E in panel (g) represent locations of observed and synthesized Stokes 
      profiles of the \ionn{He}{i}\, triplet plotted in Fig.~\ref{fits}.
      }
      \label{par_map_roi}
\end{figure}
 
\begin{figure}
\centering
      \includegraphics[width = 0.47\textwidth]{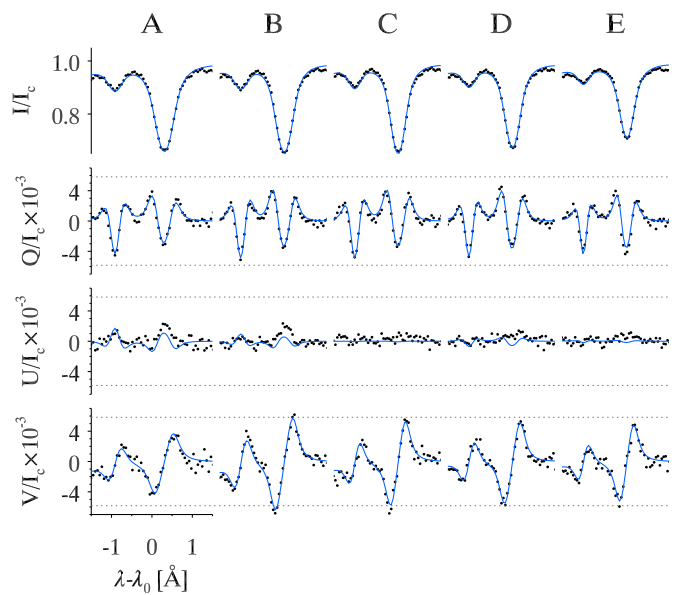}
      \caption{Observed (black dots) and synthesized (blue curves) Stokes profiles. The computed, blue profiles were obtained by the
      inversions of the \ionn{He}{i}\, triplet. Positions of columns named A--E are marked in Fig.~\ref{par_map_roi}(g).}   
      \label{fits}
\end{figure} 

\section{Formation height of the \ionn{He}{i} triplet} \label{sec_3}

Stokes $V$ profiles of sunspots observed away from the disk center in a photospheric and a chromospheric spectral line 
provide a good opportunity to estimate the difference in formation heights of the lines. Line-of-sight (LOS) magnetograms of sunspots observed 
away from disk center show an apparent neutral line (reversal of polarity) typically located on the limb side of their penumbra. The exact position of 
the neutral line with respect to the center of the sunspot depends on the view angle and the orientation of the penumbral field. 
In the typical magnetic field configuration of a sunspot with field lines expanding with height, 
the Stokes $V$ profiles along the line-of-symmetry (the line joining the disk center and sunspot center) change sign at the location where the magnetic field is perpendicular to the LOS. 
The \ionn{He}{i}\, triplet forming higher in the atmosphere, has its 
neutral line shifted towards the limb compared to that for the lower forming \ionn{Si}{i}\,10827.1\,\AA\, line. 
Knowing the separation between the neutral lines, $\Delta r$, and the viewing angle, $\theta$, one can obtain a rough estimate of the 
difference in the formation heights, $\Delta h$, using the following formula, 

\begin{center}
\begin{equation} \label{eqn_1}
\centering
\Delta h = \frac{\Delta r}{\tan \theta},
\end{equation}
\end{center}

\noindent
under the assumption that the polarity inversion line lies at about the same location when projected on the solar
surface at both heights. This is the case if the magnetic field (after spatially averaging away the fine structure)
at both heights has roughly the same inclination to the vertical (see discussion below).

Figure~\ref{prof} shows Stokes $V$ profiles both for the \ionn{Si}{i}\,10827.1\,\AA\, line and the \ionn{He}{i}\, 
triplet along the line-of-symmetry crossing the neutral line. The bottom-most and top-most profiles are located
towards the sunspot center and limb, respectively. Profiles colored in blue show the location
of the two nearest Stokes $V$ profiles with opposite sign, whereas profiles in red color identify the location
of the polarity inversion lines. It is evident that the polarity inversion line for the \ionn{He}{i}\, triplet appears 
closer to the limb compared to that for the \ionn{Si}{i}\,10827.1\,\AA\, line. The separation between the neutral
lines is $\sim1\arcsecc62$, which is equal to $\sim1530~\rm{km}$ after correcting for the foreshortening. Therefore, 
according to Eq.~\ref{eqn_1}, the difference in formation height should be around 1900~km, but Eq.~\ref{eqn_1}
is valid only when the magnetic field orientation with respect to the LOS is the same at both heights. 
The inclination of the magnetic field obtained from the \ionn{He}{i}\, triplet, however, is 
on average $12^\circ$ more vertical compared to the inclination in the photosphere \citep[see][and Sect.~\ref{sec_5}
of this article]{Joshi_2014,Joshi_2016b}. After including (adding to $\theta$) this difference of the inclination angle 
into the Eq.~\ref{eqn_1}, the difference between the formation height of the \ionn{He}{i}\, triplet and \ionn{Si}{i}\,
line is reduced to around 1250~km. Note that the formation height derived here is specific to the sunspot penumbra 
and may differ significantly in other solar features.

The \ionn{Si}{i}\,10827.1\,\AA\, line has multi-lobe Stokes $V$ profiles at and close to the periphery of the sunspot 
($r = 4\arcsecc5-5\arcsecc5$), similar to those close to the polarity inversion line ($r = 1\arcsecc5$). But, 
the multi-lobe Stokes $V$ profiles at the periphery of the sunspot are most probably due to 
unresolved mixed polarities or due to a steep gradient in the inclination of the magnetic field with height
just outside the sunspot. The He I triplet displays normal Stokes $V$ profiles at these locations, 
indicating that any disturbance to the field is restricted to layers below its formation height.

\section{Results} \label{sec_5}

Maps of the magnetic field vector and LOS-velocity obtained for the photosphere and upper chromosphere are displayed
in Fig.~\ref{par_map}. The first three maps in each panel correspond to parameters at $\log \tau = 0.0$,  $\log \tau = -0.7$ 
and $\log \tau = -2.3$ obtained by SPINOR inversions of the \ionn{Si}{i}\,10827.1\,\AA\, and \ionn{Ca}{i}\,10833.4\,\AA\,
lines. The fourth map in each panel is for the parameters returned by HeLIx$^{\rm{+}}$ inversions of the 
\ionn{He}{i}\,10830\,\AA\, triplet. The presented magnetic field vectors are transformed from the LOS frame of reference
to the solar frame of reference using a transformation matrix provided by \citet{Wilkinson_1989}. Before the transformation, the 
$180^\circ$ ambiguity in the azimuth direction of the magnetic field was resolved using the \enquote{acute angle} 
method \citep{Sakurai_1985,Cuperman_1992}. The LOS-velocities are calibrated by setting the average umbral velocity at 
$\log \tau = 0$ to zero. 

All three nodes in the photosphere show the same variation in the magnetic field strength in the azimuthal direction. 
These alternating weak and strong field structures are aligned nearly in the radial direction. In a similar way, the inclination of the magnetic
field is structured, where the relatively horizontal and vertical fields are organized in the same filamentary structure. The upper
chromosphere magnetic field strength does not display any variation in azimuthal direction, but the inclination of the field
does exhibit such a variation in the azimuthal direction which is qualitatively similar to that in the photosphere. 
% The azimuthal direction of the magnetic field is qualitatively similar at photospheric and upper chromospheric heights. 
The LOS velocities in the
photosphere show clear patterns of the radially outward directed Evershed flow, with decreasing velocities from $\log \tau = 0.0$ to $\log \tau = -0.7$, 
and velocities close to zero at  $\log \tau = -2.3$. In the upper chromosphere, the LOS-velocities have 
the opposite sign to that in the photosphere, displaying the inverse Evershed flow.

\begin{figure*}
\centering
      \includegraphics[width = 0.85\textwidth]{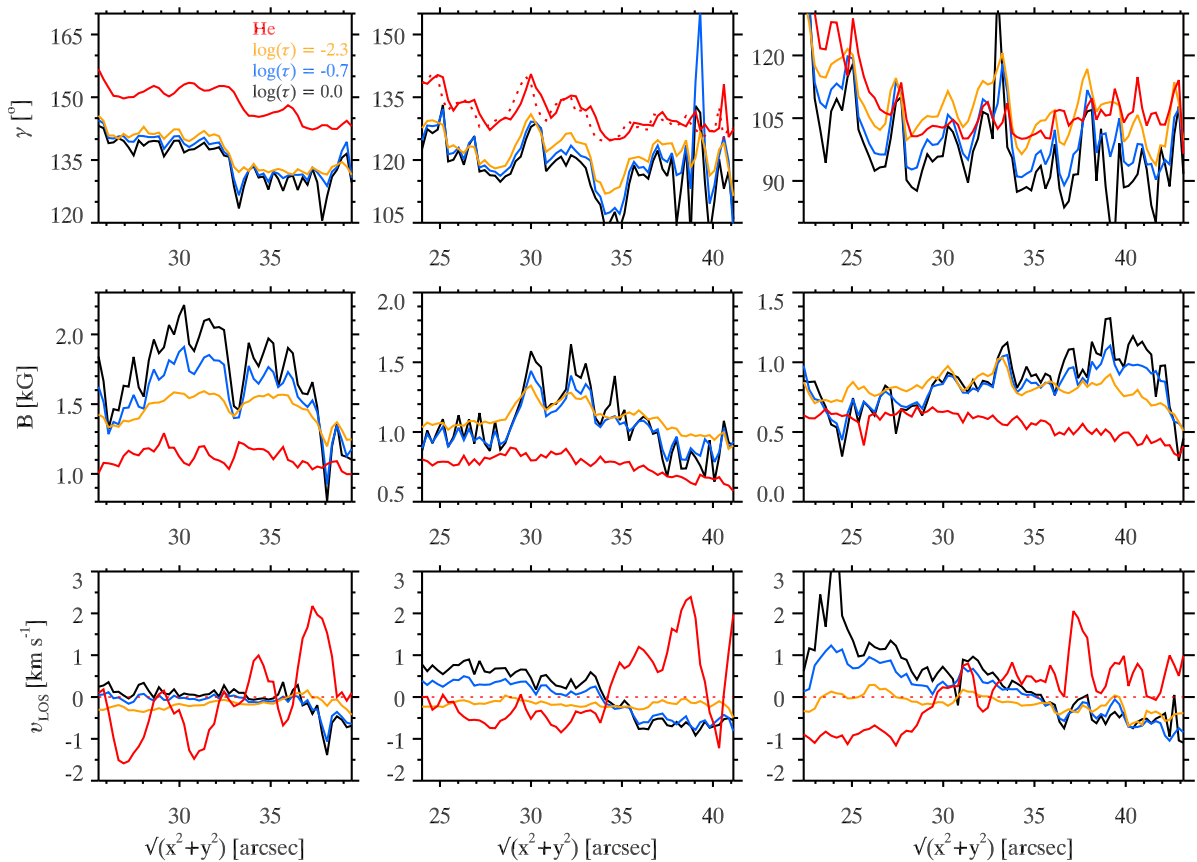}
      \caption{Profiles of the inclination of magnetic field, $\gamma$ (top row), the magnetic field strength, $B$ (middle row) and the 
      LOS-velocity, $\upsilon_{\rm LOS}$ (bottom row), along the arcs (see Fig.~\ref{FOV}(a)) at the umbra-penumbra boundary (left column), middle 
      penumbra (middle column) and outer penumbra (right column). Black, blue, yellow and red curves correspond to 
      $\log \tau= 0.0, -0.7, -2.3$, and the upper chromosphere, respectively. All the upper chromospheric
      parameters are plotted after correcting for the shift due to viewing angle and formation height. To illustrate this
      shift, an uncorrected profile is displayed by the dashed red line in the middle panel of the top row. The left of a panel corresponds to 
      the bottom of the respective arc.} 
      \label{par_prof}
\end{figure*}

\begin{figure*}[t]
\centering
      \includegraphics[width = 0.8\textwidth]{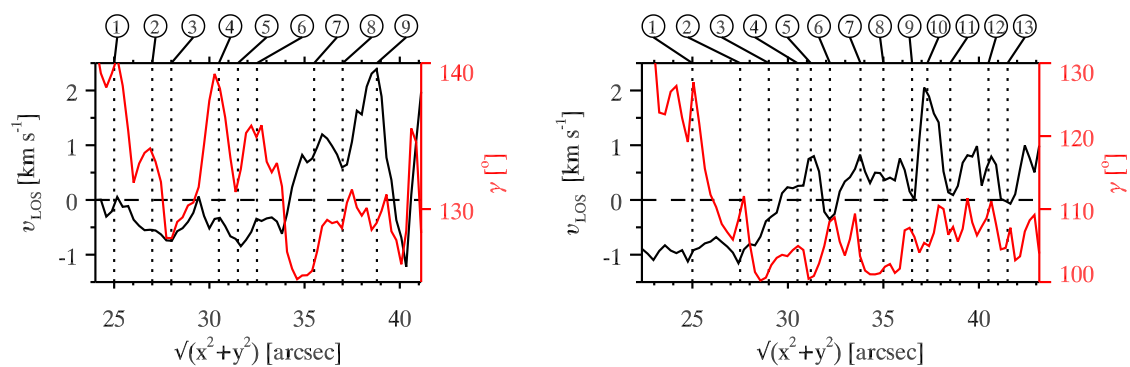}
      \caption{Variations of the LOS-velocity, $\upsilon_{\rm LOS}$ (black curve), and inclination of the magnetic field 
      $\gamma$ (red curve) obtained from the \ionn{He}{i}\, triplet, along middle 
      (left panel) and outer (right panel) arcs shown in Fig.~\ref{FOV}(a). Numbered vertical lines in both panels represent the locations of which
      variations of $\upsilon_{\rm LOS}$ and $\gamma$ are discussed in Sect.~\ref{sec_5}.}    
      \label{inc_vel}
\end{figure*}

A blow-up of the upper chromospheric magnetic field and LOS-velocity maps corresponding to the location indicated by a box in 
Fig.~\ref{par_map}(b) is presented in Fig.~\ref{par_map_roi}. The filamentary structure is apparent in the inclination, 
but not in the magnetic field strength. Contours of the inclination are plotted over the maps of $I_{\rm{c}}$, $(I_{\rm c} - I_{\rm He})/I_{\rm{c}}$,
$B$, $\phi$ and $\upsilon_{\rm{LOS}}$ in panels (a)--(d) and (f) of Fig.~\ref{par_map_roi}, respectively. These contours 
suggest that the small-scale structures in the inclination map do not have a one-to-one correspondence with any of these
parameters.

To check the reliability of the small-scale filamentary structure found in the inclination at upper chromospheric heights from 
the inversions of the \ionn{He}{i}\,10830\,\AA\, triplet, it is important to evaluate the quality of the fits to the observed
Stokes profiles. Figure~\ref{par_map_roi}(g) depicts the variation in the inclination of the magnetic field along a 
line plotted in panel (d) of Fig.~\ref{par_map_roi}. To demonstrate the quality of the fits, we choose multiple locations (marked as A-E in Fig.~\ref{par_map_roi}(g)) 
along this line where the relatively horizontal and vertical field lines appear close to each other and present the corresponding
observed and synthesized Stokes profiles in Fig.~\ref{fits}. The match between the observed and synthesized Stokes profiles is
exceptionally good for all locations. The variation in the amplitudes of the Stokes profiles also
support the variation in the inclination obtained through the inversions. For example, locations-A and -B have similar amplitudes
in Stokes $Q$/$U$, but the Stokes $V$ amplitude at location-B is larger than that at location-A, suggesting a more vertical
field at location-B, similar to the results of the inversions.

For a qualitative comparison between the photospheric and upper chromospheric magnetic field and the LOS-velocity we depict
the parameters in Fig.~\ref{par_prof} along three arcs (see Fig.~\ref{FOV}(a)) located at the umbra-penumbra boundary, the middle and 
outer penumbra, respectively.  Due to the viewing angle (away from disk center) and the difference in the formation height (1250~km) between 
the \ionn{Si}{i} line and the \ionn{He}{i} triplet, parameter maps in the upper chromosphere are shifted by projection towards the limb 
compared to those in the photosphere. The profiles of the atmospheric parameters plotted in Fig.~\ref{par_prof} are corrected for this shift. 
The inclination of the magnetic field along the arc at the umbra-penumbra boundary does not show much variation, both in the upper
chromosphere and photosphere. In the middle and outer penumbra, a variation in the inclination angle 
due to spines (vertical field) and inter-spines (horizontal field) is seen at all plotted photospheric heights. These variations in the photospheric inclination
coincide with that in the upper chromosphere. Peak-to-peak variation in the inclination decreases from $20^\circ$--$25^\circ$ at $\log \tau = 0.0$ to $\log \tau =-2.3$ 
and reach their lowest value of $10^\circ$--$15^\circ$ in the upper chromosphere. At a few locations in the outermost arc, the
magnetic field has an inclination less than $90^\circ$ at $\log \tau = 0.0$, i.e., these locations have opposite polarity compared
to the umbra. Overall the magnetic field becomes more vertical with height in the photosphere, and in the upper chromosphere 
it becomes even more vertical by an average angle of $12^\circ$ compared to that at $\log \tau = 0.0$.

The magnetic field strength shows variations due to spines and inter-spines at $\log \tau = 0.0$ along all the arcs in the penumbra.
These variations in the field strength decrease at $\log \tau = -0.7$ and  $\log \tau = -2.3$ and 
completely vanish at the upper chromosphere. Some parts (mostly at inter-spines and at their edges) of the arcs in the middle and outer penumbra have a higher
magnetic field strength at $\log \tau=-0.7$ and $\log \tau=-2.3$ than at $\log \tau=0.0$.

The gas in the arc at the umbra-penumbra 
boundary is nearly at rest along the LOS at all the photospheric heights, while in the upper chromosphere, the velocity fluctuate strongly, with multiple locations of
upflows and downflows (also see Fig.~\ref{par_map}). The middle and outer arc indicate small-scale variations in the inverse Evershed
flow at chromospheric height. To investigate the relation between these small-scale variations and the inclination angle of the magnetic field, if there 
is any, we plot both quantities along the middle and outer arc in Fig.~\ref{inc_vel}. Both panels in ~Fig.~\ref{inc_vel}
suggest that the inverse Evershed flow and the inclination have variations on similar spatial scales. There are some hints that the
stronger flows corresponds to the more horizontal magnetic field and the more vertical field has weaker flows (see for example $1^{\rm{st}}$, 
$3^{\rm{rd}} - 7^{\rm{th}}$ and $8^{\rm{th}}$ dotted vertical line in the left panel and  $5^{\rm{th}} - 9^{\rm{th}}$ line in the 
panel on the right in ~Fig.~\ref{inc_vel}). However, this relation between the inclination of the field and inverse Evershed 
flow is not particularly robust and is not present at all locations.

\section{Summary and Discussions} \label{sec_6}

We have presented the magnetic field structure of a sunspot penumbra in the upper chromosphere at high spatial resolution and have compared 
it with that in photospheric layers. In the photospheric layers the well-known spine and inter-spine structure of penumbrae in the inclination of the magnetic field are clearly visible in our data
\citep{Lites_1993,Title_1993,Solanki_1993a,Martinez_2000a,Langhans_2005,Scharmer_2013,Tiwari_2013}. Our analysis reveals that the 
spine and inter-spine structure is more prominent at $\log \tau = 0$ than at  $\log \tau = -0.7, -2.3$, in agreement 
with results of \citet{Borrero_2008}, \citet{Scharmer_2013} and \citet{Tiwari_2013}. The striking result obtained here are the 
observation of variations in the magnetic field inclination in the upper chromosphere. 
Such variations are most prominent along a cut in the azimuthal direction around the sunspot, and they resemble the 
spine and inter-spine structure in the photosphere. Our analysis reveals that these variations in the upper chromosphere coincide remarkably well with  
those in the photosphere, with typical peak to peak variations in the inclination of the field 
of $10^\circ$--$15^\circ$, compared to $20^\circ$--$25^\circ$ at $\log \tau = 0$. 

A radially aligned filamentary structure in the magnetic field strength is also visible in our results at all photospheric layers
with decreasing variation with height. The upper chromospheric map
of magnetic field strength does not show any significant variation in the lateral direction, but only a monotonic decrease 
from the inner to outer penumbra \citep[also see,][]{Joshi_2014, Schad_2015}. The absence of 
fine structure in the magnetic field strength map of the penumbra at the upper chromosphere is probably due to the dominance of magnetic 
pressure over gas pressure at this height and the absence of significant turbulence, contrary to the turbulent photosphere with 
high plasma-$\beta$ (ratio of plasma pressure to magnetic pressure), where the magnetic field gets structured on fine spatial scale. 
The presence of spine and inter-spine structure found in the maps of magnetic field inclination in the upper chromosphere suggest 
that the orientation of field lines at the photospheric heights is preserved by some extend up to $\sim$1250~km above the 
formation height of the \ionn{Si}{i}\,10827.1\,\AA\, line.

The observed variations in the magnetic field inclination in the lateral direction might be underestimated 
at the upper chromospheric height due to the limited resolution of our data. 
This speculation arises from the fact that our results do not show opposite polarity (compared to 
magnetic polarity in the umbra) lanes as observed by \citet{Scharmer_2013}, \citet{Ruiz_2013} and \citet{Tiwari_2013} at the 
edges of penumbral filaments in the photosphere. This is because our observations have a lower resolution compared
to the observations analyzed in the above studies. Also, overlooking of the spatial PSF and straylight may contribute to underestimating
in the variations of the inclination of the magnetic field. However, we do observe some opposite polarity patches at the tails of 
penumbral filaments at $\log \tau$ = 0, similar to those observed by \citet{Ruiz_2013}, \citet{Franz_2013}, \citet{Scharmer_2013}
and \citet{Tiwari_2013}. Finally, our study also confirms the localized presence of a negative vertical gradient 
of the magnetic field strength in the photospheric layers of the penumbra, i.e., magnetic field strength decreases with depth as reported by 
\citet{Tiwari_2015} and \citet{Joshi_2016a}. The latter authors interpreted the observed negative vertical gradient as a result of 
the strong magnetic field from spines closing above the weaker field inter-spines and 
also as a result of unresolved opposite polarity patches in the deep photosphere.

The magnetic field strength, in general, decreases with height in the photosphere and
into the upper chromosphere. The average magnetic field strength close to the inner and outer penumbra is around 1.7~kG and 
0.8~kG in the photosphere ($\log \tau = 0.0$), respectively, and  1.05~kG and 0.55~kG in the upper chromosphere. 
If we use the difference between the formation height of the \ionn{Si}{i}\,10827.1\,\AA\, line and the \ionn{He}{i}\, triplet of 1250~km that we have determined, 
the vertical gradient turns out to be $\sim0.6$\,G\,km$^{-1}$ and $\sim0.2$\,G\,km$^{-1}$ in the inner and outer parts of the penumbra, respectively.  
The difference in the formation height of the \ionn{Si}{i}\, line and the \ionn{He}{i}\, triplet is estimated using the separation between the polarity-inversion lines displayed by these spectral lines. 
These values of the vertical gradient of the magnetic field strength are similar to those reported earlier using the same spectral lines \citep[see][]{Rueedi_1995b,Joshi_2014,Schad_2015,Joshi_2016b}.
Our results indicate that in general the magnetic field of the penumbra becomes more vertical with height in the photosphere 
\citep[also see][]{Borrero_2011,Tiwari_2013} and in the upper chromosphere it is more vertical by $\sim 12^\circ$ compared to that
in the photosphere ($\log \tau = 0.0$). This difference in the magnetic field inclination between the photosphere and upper chromosphere
is similar to that found by \citet{Joshi_2014} and \citet{Joshi_2016b}.

In summary, we report, to our knowledge, the first observations of fine structure of the penumbral magnetic 
field of a sunspot in the upper chromosphere, which displays lateral variation in the magnetic field inclination 
resembling the spine and inter-spine structure in the photosphere. The observed inhomogeneity in the inclination 
of the chromospheric magnetic field on small-scales may play a role in driving small-scale chromospheric and transition region dynamics,
such as, penumbral micro-jets \citep{2007Sci...318.1594K, 2008A&A...488L..33J, 2013ApJ...779..143R, 2015ApJ...811L..33V, 
2016ApJ...816...92T} and transient bright dots \citep{2014ApJ...790L..29T}. Scrutiny of this scenario requires a detailed analysis of the 
dynamics of the penumbral magnetic field in the chromosphere, and, of its coupling to the photosphere. A possible 
observational setup for this purpose would be similar to that presented in the current article but either in a  
sit-and-stare mode or with repetitive small raster scans (about $3\varcsec-4\varcsec$ wide) parallel to penumbral filaments.
These would be accompanied by images in \ionn{Ca}{ii}\,H or K and Interface Region Imaging Spectrograph \citep[IRIS,][]{2014SoPh..289.2733D} data.   

\begin{acknowledgements}

The 1.5-meter GREGOR solar telescope was built by a German consortium under the leadership of the Kiepenheuer-Institut 
f\"ur Sonnenphysik in Freiburg with the Leibniz-Institut f\"ur Astrophysik Potsdam, the Institut f\"ur Astrophysik G\"ottingen, 
and the Max-Planck-Institut f\"ur Sonnensystemforschung in G\"ottingen as partners, and with contributions by the 
Instituto de Astrof\'isica de Canarias and the Astronomical Institute of the Academy of Sciences of the Czech Republic.
JJ acknowledges support from the CHROMOBS project funded by the Kunt and Alice Wallenberg Foundation.
This work was partly supported by the BK21 plus program through the National Research Foundation (NRF) funded by 
the Ministry of Education of Korea.
This study is supported by the European Commissions FP7 Capacities Programme under the Grant Agreement number 312495.
The GRIS instrument was developed thanks to the support by the Spanish Ministry of Economy and Competitiveness through 
the project AYA2010-18029 (Solar Magnetism and Astrophysical Spectropolarimetry).

\end{acknowledgements}

\bibliography{joshi_gris_arxiv}

\end{document}